\documentclass[10pt, a4paper]{article}

\usepackage{graphicx}
\usepackage{amssymb,amsmath}
\usepackage{a4wide}

\newtheorem{thm}{Theorem}
\begin{document}

\title{The Multiple Access Channel with Feedback and Correlated Sources}

\author{Lawrence Ong and Mehul Motani\\
Department of Electrical and Computer Engineering\\
National University of Singapore\\
Email: \{lawrence.ong, motani\}@nus.edu.sg\\}
\date{}

\maketitle

\begin{abstract}
In this paper, we investigate communication strategies for the multiple access channel with feedback and correlated sources (MACFCS). The MACFCS models a wireless sensor network scenario in which sensors distributed throughout an arbitrary random field collect correlated measurements and transmit them to a common sink.
We derive achievable rate regions for the three-node MACFCS. First, we study the strategy when source coding and channel coding are combined, which we term \emph{full decoding at sources}. Second, we look at several strategies when source coding and channel coding are separated, which we term \emph{full decoding at destination}. From numerical computations on Gaussian channels, we see that different strategies perform better under certain source correlations and channel setups.
\end{abstract}

\section{Introduction}

In this paper, we investigate the the \emph{multiple access channel with feedback and correlated sources} (MACFCS),
as defined in \cite{ongmotani05c}.
This channel combines the multiple access channel with correlated sources (MACCS) and the multiple access channel with feedback (MACF). It consists of multiple sources which send possibly correlated data to a single destination. At the same time, each source receives feedback from the channel.

The MACCS (with a common part) was studied by Slepian and Wolf~\cite{slepianwolf73b}, who derived an achievable rate region. In their paper, separate source coding and channel coding are used, where source coding is first performed to remove the correlation between the two sources and then channel coding for the multiple access channel (MAC) with independent sources is employed. The MACCS (with possibly no common part) was considered by Cover \emph{et al.}~\cite{covergamal80}. They showed, by using a simple example, that separating source and channel coding is not optimal and derived an achievable rate region for the MACCS.

The MACF (with independent sources) was investigated by Cover and Leung~\cite{coverleung81}. In their model, there are two sources and all nodes, i.e., the two sources and the destination, receive the same channel output. A year later, Carleial~\cite{carleial82} further generalized the channel to the case where each node receives a different channel output signal.

\begin{figure}[t]
    \centering
        \includegraphics[width=8cm]{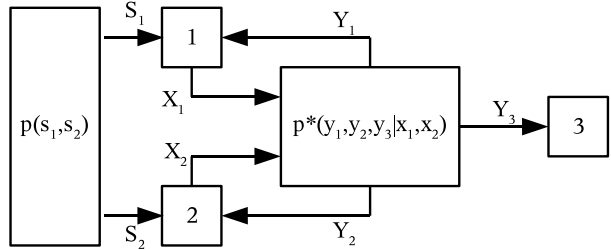}
    \caption{The 3-node multiple access channel with feedback and correlated sources.}
    \label{fig:macfcs}
\end{figure}

\subsection{Channel Model}
Combining the MACF and the MACCS, we get the MACFCS, depicted in Figure~\ref{fig:macfcs}. The three-node discrete memoryless MACFCS is denoted by $\big( \mathcal{S}_1 \times \mathcal{S}_2, p(s_1,s_2), \mathcal{X}_1 \times \mathcal{X}_2, p^*(y_1,y_2,y_3 |x_1,x_2), \mathcal{Y}_1 \times \mathcal{Y}_2  \times \mathcal{Y}_3 \big)$. $s_1 \in \mathcal{S}_1$ and $s_2 \in \mathcal{S}_2$ are the source messages to nodes 1 and 2 respectively and they are drawn from the discrete bivariate distribution $p(s_1,s_2)$. Here, $\mathcal{S}_1, \mathcal{S}_2, \mathcal{X}_1, \mathcal{X}_2, \mathcal{Y}_1, \mathcal{Y}_2,$ and $\mathcal{Y}_3$ are seven finite sets. $p^*(y_1,y_2,y_3 | x_1,x_2)$ defines the channel transition probability on $\mathcal{Y}_1 \times \mathcal{Y}_2 \times \mathcal{Y}_3$ for each $(x_1,x_2) \in \mathcal{X}_1 \times \mathcal{X}_2$. $x_1$ and $x_2$ are the inputs to the channel from nodes 1 and 2 respectively. $y_1$, $y_2$, and $y_3$ are the channel outputs to nodes 1, 2, and the destination respectively. We say that the random variable pair $(S_1,S_2)$ can be reliably transmitted to the destination if the probability that the destination wrongly decodes a pair of $(\mathbf{S}_1,\mathbf{S}_2) \in \mathcal{S}_1^n \times \mathcal{S}_2^n$ in each $n$ channel uses can be made arbitrarily small, for large $n$.

\subsection{The Gaussian MACFCS} \label{sec:macfcs_gaussian}
In the $T$-node Gaussian MACFCS (G-MACFCS), the channel outputs are given by
\begin{equation}
Y_t = \sum_{\substack{i=1\\i \neq t}}^{T-1} \sqrt{\kappa d_{it}^{-\eta}}X_i + Z_t, \quad\quad t=1,2,\dotsc,T,
\end{equation}
where $Y_t$ is the received signal at node $t$. $X_i$ is the signal transmitted by node $i$, with power constraints $E[X_i^2] \leq P_i$. $Z_t$ is an independent, zero-mean, Gaussian random variable at the receiver of node $t$ with variance $N_t$. $\kappa$ is a positive constant and $\eta$ is the path loss exponent. $d_{it}$ is the distance between nodes $i$ and $t$. We assume perfect echo cancellation, meaning that a node can cancel its own transmission perfectly.

Murugan \emph{et al.}~\cite{murugangopala04} looked at a similar problem, assuming Gaussian channels and a total average power constraint on both sources. Their coding approach is based on joint source-channel coding using time division multiple access (TDMA) with the nodes operating in half-duplex. Our work differs from \cite{murugangopala04} in that we consider general channels (including discrete memoryless and Gaussian channels) with  full-duplex nodes, in which the source nodes can transmit and receive simultaneously.  For the Gaussian case, we impose individual power constraints on the sources. 
We consider both joint and separate source-channel coding strategies.

In brief, we classify coding strategies for the MACFCS into two categories: full decoding at sources and full decoding at the destination. We describe these strategies in more detail in the following sections. The rest of the paper is organized as follows. In Section~\ref{sec:df}, we derive achievable rate regions for the MACFCS using a full decoding at sources strategy. In Section~\ref{sec:fdad}, we derive achievable rate regions for the MACFCS using full decoding at destination strategies. We compare the performance of the different strategies in Section~\ref{sec:numerical}.

\section{Full Decoding at Sources}\label{sec:df}

For full decoding at sources, the general idea is for the sources to communicate so that each source has full information about what the other sources have. They then cooperate to send the signal \emph{coherently} to the destination. Each source node transmits cooperative information of the previous block (which it decodes from other nodes) and new information (which is to be decoded by other sources and the destination) in one block.

We consider the following correlation structure in which the sources have a common part to send. Let $I, J,$ and $K$ be three independent random variables. Node 1 receives $S_1 = (I,J)$ and node 2 receives $S_2 = (I,K)$.

We note that node 1 does not know $K$ and node 2 does not know $J$. The idea here is for node 1 to decode $K$ (from node 2's transmission) and for node 2 to decode $J$. After decoding, both nodes have the full information $(I,J,K)$. Each node then sends the fully decoded information as well as its new received information that is unknown to and to be decoded by other nodes. In summary, node 1 sends $(J, I', J', K')$ and node 2 sends $(K, I', J', K')$, where prime denotes the previous block's information. Using this strategy, we have the following theorem.

\begin{thm}\label{thm:achievability_macfcs_df}
In a three-node discrete memoryless MACFCS, the source messages $(S_1,S_2)$ can be reliably transmitted to the destination if the following hold:
\begin{subequations}
\begin{align}
&H(S_1,S_2) < I(X_1,X_2;Y_3|Q), \nonumber \\
& H(S_1|S_2)  < \min \Big[ I(X_1;Y_2|Q,W_0,W_1,W_2,X_2), I(W_1;Y_3|Q,W_0,W_2) + I(X_1;Y_3|Q,W_0,W_1,W_2,X_2) \Big], \nonumber \\
& H(S_2|S_1) < \min \Big[ I(X_2;Y_1|Q,W_0,W_1,W_2,X_1), I(W_2;Y_3|Q,W_0,W_1) + I(X_2;Y_3|Q,W_0,W_1,W_2,X_1) \Big], \nonumber \\
&I(S_1;S_2) < I(W_0;Y_3|Q,W_1,W_2), \nonumber \\
&H(S_1) < I(W_0,W_1;Y_3|Q,W_2) + I(X_1;Y_3|Q,W_0,W_1,W_2,X_2), \nonumber \\
&H(S_2) < I(W_0,W_2;Y_3|Q,W_1)  + I(X_2;Y_3|Q,W_0,W_1,W_2,X_1), \nonumber \\
&H(S_1|S_2) + H(S_2|S_1) < I(W_1,W_2;Y_3|Q,W_0)  + I(X_1,X_2;Y_3|Q,W_0,W_1,W_2), \nonumber
\end{align}
\end{subequations}
where $p(q,x_1,x_2,y_1,y_2,y_3,w_0,w_1,w_2) = p(q) p (w_0|q)p(w_1|q)p(w_2|q)p(x_1|q,w_0,w_1,w_2)$\\$p(x_2|q,w_0,w_1,w_2)p^*(y_1,y_2,y_3|x_1,x_2,x_3)$. $W_0$, $W_1$ and $W_2$ are auxiliary random variables. $Q$ is the time sharing variable.
\end{thm}

We note that by setting
\begin{subequations}
\begin{align}
& Q=0: W_0=0, W_1=0, W_2=0, X_2=0, \\
& Q=1: W_0=0, W_1=0, W_2=0, X_1=0, \\
& Q=2: X_1 = f(W_0,W_1,W_2), X_2 = f(W_0,W_1,W_2),
\end{align}
\end{subequations}
for some deterministic function $f(\cdot)$, we end up with the half-duplex coding scheme in \cite{murugangopala04}. In the decoding in \cite{murugangopala04}, the destination only decodes at time $Q=2$. In other words, they exclude the terms $I(\dotsm; Y_3|Q=i,\dotsm)$ for $i=1,2$. In Theorem~\ref{thm:achievability_macfcs_df}, the destination decodes for all $Q=0,1,2$ and hence the rate achievable can be higher.

\begin{figure}[t]
    \centering
        \includegraphics[width=9cm]{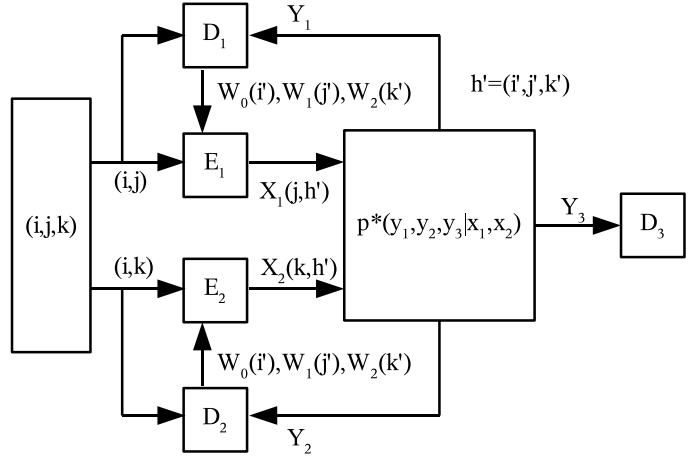}
    \caption{Coding for the multiple access channel with feedback and correlated sources using the decode-forward strategy.}
    \label{fig:macfcs_dec_fwd}
\end{figure}

\emph{Sketch of proof for Theorem~\ref{thm:achievability_macfcs_df}:} We give an outline of the proof for Theorem~\ref{thm:achievability_macfcs_df}. We ignore $Q$ in the following discussion to simply the expressions. Using Slepian and Wolf's Theorem 2 in \cite{slepianwolf73}, when node 1 only knows $S_1=(I,J)$ and node 2 knows $S_2=(I,K)$, node 1 can encode $J$ using $H(S_1|S_2)$ bits and it can decoded by node 2. Similarly, node 2 can use $H(S_2|S_1)$ bits to encode $K$.

Now, we explain the auxiliary random variables used. Referring to Fig.~\ref{fig:macfcs_dec_fwd}, $W_0$ carries the common information from the previous block,  $W_1$ carries node 1's private information from the previous block, while $W_2$ carries node 2's private information from the previous block. In each block of transmission, node 1 needs to decode node 2's new private information. This can be done reliably if
$H(S_2|S_1) < I(X_2;Y_1|W_0,W_1,W_2,X_1)$.
Similar, node 2 needs to decode node 1's private information and this can be done reliably if
$ H(S_1|S_2) < I(X_1;Y_2|W_0,W_1,W_2,X_2)$.

At the destination, the common information can be reliably decoded if
$I(S_1;S_2) < I(W_0;Y_3|W_1,W_2)$.
Node 1's private information can be reliably decoded over two blocks if
$H(S_1|S_2) < I(W_1;Y_3|W_0,W_2) + I(X_1;Y_3|W_0,W_1,W_2,X_2)$.
Similarly, node 2's private information can be reliably decoded over two blocks at the destination if
$H(S_2|S_1) < I(W_2;Y_3|W_0,W_1) + I(X_2;Y_3|W_0,W_1,W_2,X_1)$.
To ensure that node 1's message, node 2's message, nodes 1 and 2's common parts, and nodes 1 and 2's messages can be reliably decoded at the destination, the following need to be satisfied respectively.
\begin{subequations}
\begin{align}
& H(S_1) < I(W_0,W_1;Y_3|W_2) + I(X_1;Y_3|W_0,W_1,W_2,X_2), \nonumber\\
& H(S_2) < I(W_0,W_2;Y_3|W_1) + I(X_2;Y_3|W_0,W_1,W_2,X_1), \nonumber\\
& H(S_1|S_2) + H(S_2|S_1) < I(W_1,W_2;Y_3|W_0)  + I(X_1,X_2;Y_3|W_0,W_1,W_2),\nonumber\\
& H(S_1,S_2) < I(X_1,X_2;Y_3)\nonumber.
\end{align}
\end{subequations}
The total probability of error can be bounded for large $n$ if the equations above hold. Hence, we have Theorem~\ref{thm:achievability_macfcs_df}.

We note that in our derivation, we use a correlation structure with an explicit common part for clearer illustration. However, the analysis can be generalized to the case where there is no common part, and hence Theorem~\ref{thm:achievability_macfcs_df} can be proven for sources with any arbitrary correlation structure.


\section{Full Decoding at Destination}\label{sec:fdad}
For full decoding at the destination, source coding is first performed at every source node using the technique described in \cite{slepianwolf73b}. This removes the correlation among the sources.

At this point, we have turned the problem into channel coding for the MACF with independent sources. An achievable region of the MACF was obtained by Carleial~\cite{carleial82}. We call that the partial decode-forward strategy. In this paper, we find another achievable region for the MACF using the compress-forward strategy (see Section~\ref{sec:cf}). Combining the rate constraints of the source coding (for correlated sources) and the channel coding (for the MACF), we arrive at other achievable rate regions for the MACFCS. To the best of our knowledge, the compress-forward strategy has not been studied on the MACF. Ignoring the feedback of the MACF, we can combine the source coding with correlated source and channel coding for the MAC and arrive at yet another set of achievable rates.

\subsection{Source Coding for Correlated Sources} \label{sec:cs}
Source coding for correlated sources is performed prior to channel coding for MACF. First, we consider a noiseless channel. With node 1 knowing only $S_1$, node 2 knowing only $S_2$, the destination can reconstruct $(S_1,S_2)$ reliably if node 1 encodes $S_1$ with rate $R_1$ and node 2 encodes $S_2$ with rate $R_2$ \cite{slepianwolf73b}, where
\begin{subequations}
\begin{align}
R_1 & \geq H(S_1|S_2), \label{eq:cs_start} \\
R_2 & \geq H(S_2|S_1), \\
R_1 + R_2 & \geq H(S_1,S_2). \label{eq:cs_end}
\end{align}
\end{subequations}

\subsection{Combining with Channel Coding for the MAC}

The simplest way of performing channel coding after source coding is to use that of the MAC. In this case, we disregard the feedback of the channel to the source nodes. Each source now simple sends independent messages as it would in the MAC. In this case, the source nodes do not cooperate. The capacity of the MAC can be found in \cite{liao72} and \cite{ahlswede74}. Combining these results and the source coding constraints, the messages $(S_1,S_2)$ of the MACFCS can be reliably transmitted to the destination if the following hold.
\begin{subequations}
\begin{align}
H(S_1|S_2) & \leq I(X_1;Y_3|X_2) , \\
H(S_2|S_1) & \leq I(X_2;Y_3|X_1), \\
H(S_1,S_2) & \leq I(X_1,X_2;Y_3),
\end{align}
\end{subequations}
where $p(x_1,x_2)=p(x_1)p(x_2)$.

\subsection{Combining with  Partial Decode-Forward for the MACF}

An achievable rate region for the MACFCS can be derived by combining the source coding rate constraints (\eqref{eq:cs_start}-\eqref{eq:cs_end} in Section~\ref{sec:cs}) and the channel coding constraints for the MACF ((3a), (3b), (7a)-(7q) in \cite{carleial82}). We call the strategy used in \cite{carleial82} the partial decode-forward strategy.  The proof that this rate region is achievable is straightforward.


\subsection{Combining with Compress-Forward for the MACF}\label{sec:cf}
\begin{figure}[t]
    \centering
        \includegraphics[width=8cm]{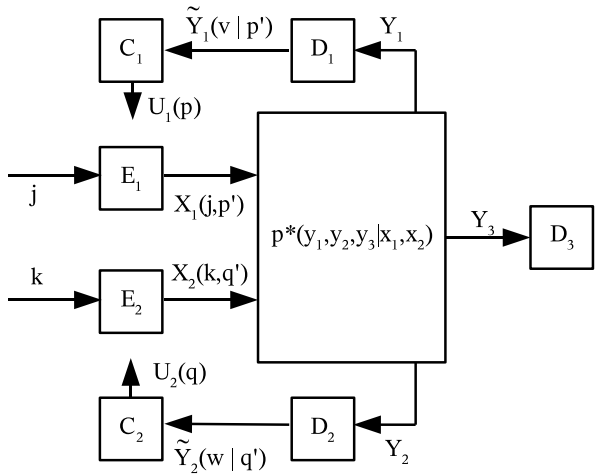}
    \caption{Coding for the multiple access channel with feedback (with independent sources) using the compress-forward strategy.}
    \label{fig:macfcs_cmp_fwd}
\end{figure}

In this section, we derive an achievable rate for the MACF using the compress-forward strategy. Combining this with the source coding rate constraints in Section~\ref{sec:cs}, we derive another achievable rate region for the MACFCS.

Using the compress-forward strategy, each node transmits independent information as well as a quantized and compressed version of its received signal. Referring to Figure~\ref{fig:macfcs_cmp_fwd}, $J$ and $K$ are independent information after source coding. Consider node 1 as an example first. From the received signal $Y_1$, it produces a quantized version $\tilde{Y}_1$. 
It then compresses $\tilde{Y}_1$ to $U_1$. In the next block, it sends new information $J$ as well as $U_1$. We can view this as node 1 helping node 2 to send a noisy, quantized, and compressed version of node 2's signal, $K$, without needing to fully decode $K$. Node 2 does likewise.

Combining the source coding the correlated sources and compress-forward channel coding, we can show that the following theorem.
\begin{thm}\label{thm:achievability_macfcs_2cf}
In a three-node discrete memoryless MACFCS. The source symbols $(S_1,S_2)$ can be reliably transmitted to the destination if
\begin{subequations}
\begin{align}
& H(S_1|S_2) < I(X_1;\tilde{Y}_1,\tilde{Y}_2,Y_3|Q,U_1,U_2,X_2), \\
& H(S_2|S_1) < I(X_2;\tilde{Y}_1,\tilde{Y}_2,Y_3|Q,U_1,U_2,X_1), \\
& H(S_1,S_2) < I(X_1, X_2; \tilde{Y}_1, \tilde{Y}_2, Y_3 |Q, U_1, U_2),
\end{align}
\end{subequations}
where the mutual information is taken over all joint probability mass functions\\ $
p(q,u_1,u_2,x_1,x_2,\tilde{y}_1,\tilde{y}_2,y_1,y_2,y_3)= p(q) p(u_1|q)p (x_1|q,u_1) p(u_2|q) p(x_2|q,u_2)  p(\tilde{y}_1|q,y_1,x_1,u_1)$\\$ p(\tilde{y}_2|q,y_2,x_2,u_2) p^*(y_1,y_2,y_3|x_1,x_2),$
subjected to the following constraints
\begin{subequations}
\begin{align}
I(U_1;Y_3|Q,U_2) & > I(\tilde{Y}_1;Y_1|Q,X_1,U_1) - I(\tilde{Y}_1;Y_3|Q,\tilde{Y}_2,U_1,U_2), \nonumber \\
I(U_2;Y_3|Q,U_1) & > I(\tilde{Y}_2;Y_2|Q,X_2,U_2) - I(\tilde{Y}_2;Y_3|Q,\tilde{Y}_1,U_1,U_2), \nonumber \\
I(U_1,U_2;Y_3|Q) & > I(\tilde{Y}_1;Y_1|Q,X_1,U_1) + I(\tilde{Y}_2;Y_2|Q,X_2,U_2)  - I(\tilde{Y}_1,\tilde{Y}_2;Y_3|Q,U_1,U_2). \nonumber
\end{align}
\end{subequations}
Here, $U_1, U_2, \tilde{Y_1}$, and $\tilde{Y}_2$ are auxiliary random variables. $Q$ is the time sharing variable. We ignore $Q$ in the following discussion to simply the expressions.
\end{thm}

\emph{Sketch of proof for Theorem~\ref{thm:achievability_macfcs_2cf}:} Now we give an outline of the proof for Theorem~\ref{thm:achievability_macfcs_2cf}. We use $R_1$, $R_2$ to represent the rate (after source coding) from nodes 1 and 2 respectively. $\tilde{R}_1$ and $\tilde{R}_2$ represent the rate of the quantized version of the received signals of nodes 1 and 2 respectively. $R_1'$ and $R_2'$ represent the compressed of the quantized signal of nodes 1 and 2 respectively.
Node 1 quantizes its received signal to $\tilde{Y}_1$. It can do this reliably if
$ \tilde{R}_1 > I(\tilde{Y}_1;Y_1|X_1,U_1)$.
Similarly for node 2 to quantizes its received signal, $\tilde{R}_2 > I(\tilde{Y}_2;Y_2|X_2,U_2)$.
Nodes 1 and 2 then compress the quantizes information and send them to the destination. The destination can decode the compressed information $(U_1,U_2)$ if
$R_1' < I(U_1;Y_3|U_2)$, $R_2' < I(U_2;Y_3|U_1)$, and $R_1' + R_2' < I(U_1,U_2;Y_3)$ hold.
After decoding the compressed information, the destination finds the quantized information $\tilde{Y}_1$ and $\tilde{Y}_2$. The following inequalities must be satisfied.
\begin{subequations}
\begin{align}
\tilde{R}_1 & < I(\tilde{Y}_1;Y_3|\tilde{Y}_2,U_1,U_2) + R_1', \\
\tilde{R}_2 & < I(\tilde{Y}_2;Y_3|\tilde{Y}_1,U_1,U_2) + R_2', \\
\tilde{R}_1 + \tilde{R}_2 & < I(\tilde{Y}_1,\tilde{Y}_2;Y_3|U_1,U_2) + R_1' + R_2'.
\end{align}
\end{subequations}
After decoding the quantized information, the destination uses $\tilde{Y}_1, \tilde{Y}_2$, and $Y_3$ to decode the source messages. This can be done reliably if
\begin{subequations}
\begin{align}
R_1 & < I(X_1;\tilde{Y}_1,\tilde{Y}_2,Y_3|U_1,U_2,X_2), \\
R_2 & < I(X_2;\tilde{Y}_1,\tilde{Y}_2,Y_3|U_1,U_2,X_1), \\
R_1 + R_2 & < I(X_1, X_2; \tilde{Y}_1, \tilde{Y}_2, Y_3 | U_1, U_2).
\end{align}
\end{subequations}

Combining these rate constraints for the MACF using the compress-forward strategy and the constraints for the source coding, \eqref{eq:cs_start}-\eqref{eq:cs_end}, we get Theorem~\ref{thm:achievability_macfcs_2cf}.

\begin{figure*}[t]
\begin{minipage}[t]{0.48\linewidth}
\centering
\includegraphics[width=0.9\textwidth]{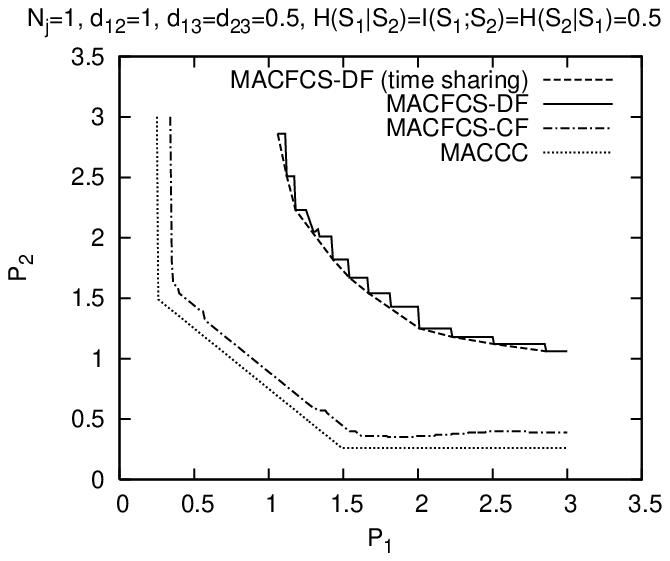}
\caption{Minimum power required to transmit $(s_1,s_2)$ to the destination per channel use, with weak inter-source link.}
\label{fig:power_sym_equidistance}
\end{minipage}
\hfill
\begin{minipage}[t]{0.48\linewidth}
\centering
\includegraphics[width=0.9\textwidth]{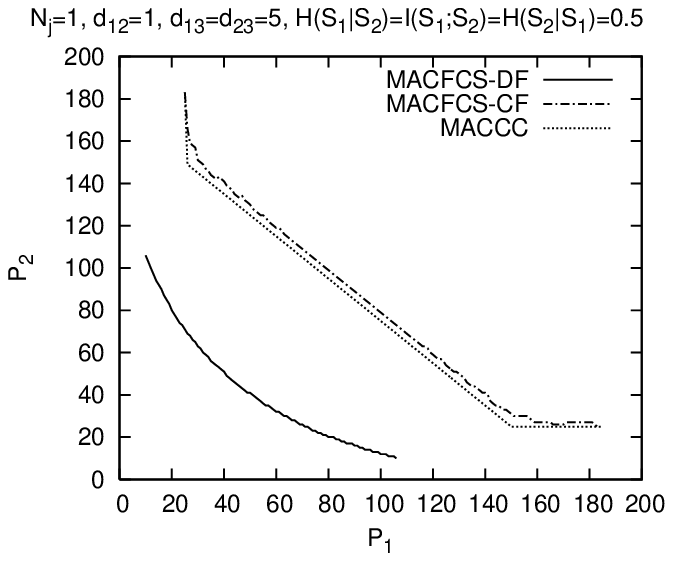}
\caption{Minimum power required to transmit $(s_1,s_2)$ to the destination per channel use, with weak source-destination link.}
\label{fig:power_large_d13}
\end{minipage}
\end{figure*}

\section{Numerical Computations}\label{sec:numerical}
In this section, we plot and compare the powers required to transmit data on the three-node G-MACFCS using different strategies. Although Gaussian channel input distributions may not be optimal, we choose $X_1$, $X_2$, and the auxiliary random variables to be Gaussian for the sake of comparison. We also assume that Theorem~\ref{thm:achievability_macfcs_2cf}, which uses the compress-forward strategy, holds for Gaussian random variables \cite[Remarks 28,30]{kramergastpar04}.

Figs.~\ref{fig:power_sym_equidistance} and \ref{fig:power_large_d13} show the minimum average transmit powers required for nodes 1 and 2 to reliably transmit a pair of discrete messages $(S_1,S_2)$ per channel use. We fix the sources to have the following structures: $H(S_1|S_2)= H(S_2|S_1)=I(S_1;S_2)=0.5$.

When the inter-source link is weak (Fig.~\ref{fig:power_sym_equidistance}), full decoding at destination with the compress-forward strategy (MACFCS-CF) and separate source-channel coding with MAC channel coding (MACCC) perform better than full decoding at sources with the decode-forward strategy (MACFCS-DF). It can be seen from the graph that less power is required using MACFCS-CF and MACCC compared to MACFCS-DF. Using MACFCS-DF, each source node needs to get full information of what other nodes transmit. This imposes an extra constraint on the transmit power at the source nodes.

When the source-destination links are weak (Fig.~\ref{fig:power_large_d13}), MACFCS-DF performs better than MACFCS-CF and MACCC. This is because the transmission bottleneck is now at the source-destination link. Coherent combining is possible using MACFCS-DF and it yields a significant gain in the rate on the source-destination link.

The staircase behavior in Fig.~\ref{fig:power_sym_equidistance} is caused by different inequalities governing the overall achievable rate in Theorem~\ref{thm:achievability_macfcs_df} and optimization of $\{\alpha_{ij}\}$. Since the region is non-convex, time sharing enlarges the region.

\begin{figure*}[t]
\begin{minipage}[t]{0.48\linewidth}
\centering
\includegraphics[width=0.95\linewidth]{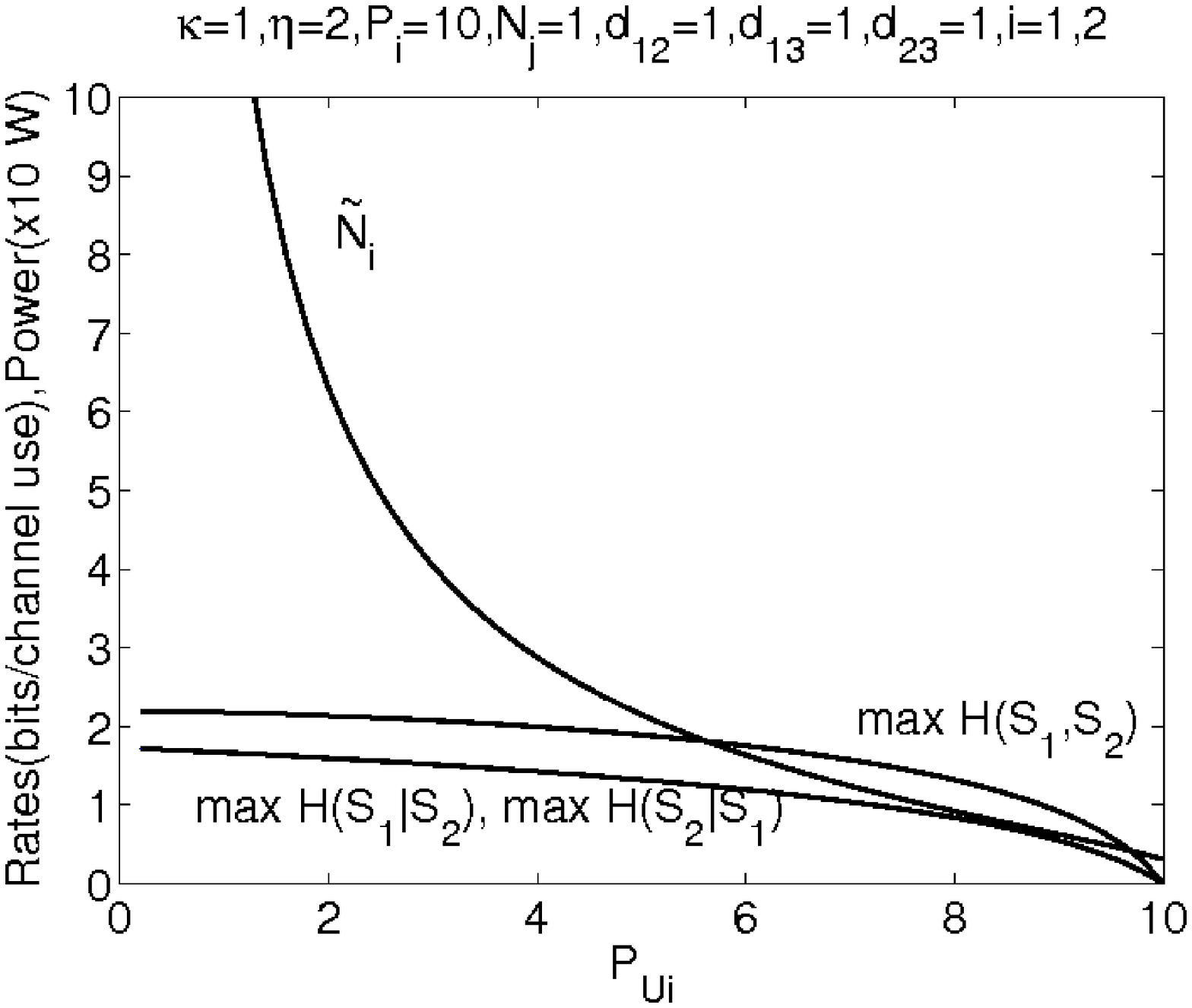}
\caption{Achievable rates of the compress-forward strategy on the three-node G-MACFCS with equal node distance and smallest feasible $\tilde{N}_1$(=$\tilde{N}_2$).}
\label{fig:rates_cf_tilde_n}
\end{minipage}
\hfill
\begin{minipage}[t]{0.48\linewidth}
\centering
\includegraphics[width=0.9\textwidth]{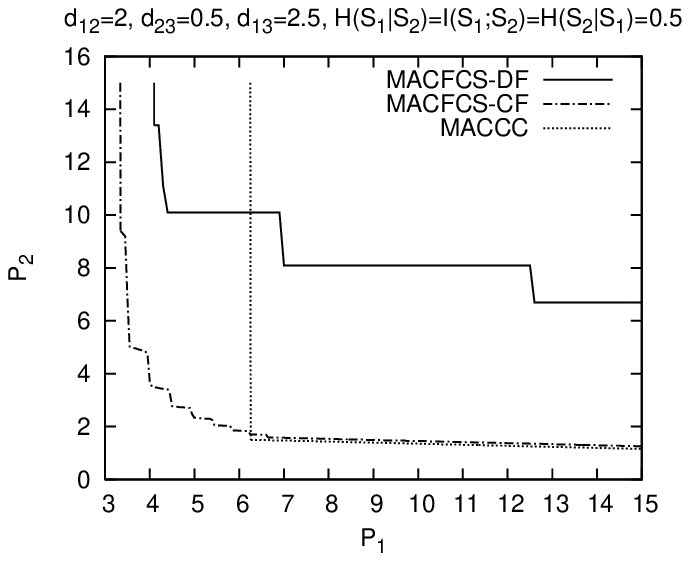}
\caption{Minimum power required to transmit $(s_1,s_2)$ to the destination per channel use, in a linear topology where node 2 is closer to the destination.}
\label{fig:rates_cfdf_linear}
\end{minipage}
\end{figure*}

We note that in Figs.~\ref{fig:power_sym_equidistance} and \ref{fig:power_large_d13}, MACCC performs slightly better than MACFCS-CF. This can be explained using Fig.~\ref{fig:rates_cf_tilde_n}, which shows the smallest feasible $\tilde{N}_1$ or $\tilde{N}_2$ required for a certain channel setup, $P_{Ui}$, and $P_{Vi}$. Recall that $P_{Ui}$ is the power allocated to transmit the compressed information (old message) to the destination while $P_{Vi}$ carries new information. We note that $P_{Ui} + P_{Vi} = 10$. From the graph, we see that when we allocate more power to $P_{Ui}$, more power is used to transmit the compressed message. In this case, the compression noise $\tilde{N}_i$ is smaller. When all transmit power is used allocated to $P_{Ui}$, no power is used to transmit new messages and hence the overall transmission rate is zero. In the figure, the rate is maximized when we set $P_{Ui}\rightarrow 0$. When we allocate all transmit power to $P_{Vi}$, we are essentially doing channel coding for the MAC. However, a large $\tilde{N}_i$ ($\rightarrow \infty$) is required when $P_{Ui} \rightarrow 0$. Hence the rate we obtain using MACFCS-CF approaches that of MACCC when $P_{Ui} \rightarrow 0$.

MACCC does not always perform better than MACFCS-CF. The latter outperforms the two other strategies when node 2 is placed closer to the destination than to node 1. This is depicted in Fig.~\ref{fig:rates_cfdf_linear}. In short, our numerical computations indicate that the performance of the different strategies varies with the channel settings.



%

\bibliography{bib}

\end{document}